\documentclass[11pt]{article}
\usepackage{graphicx} 
\usepackage{svg}
\usepackage{booktabs}
\usepackage{multirow}
\usepackage{longtable}
\usepackage{tabularx}
\usepackage{ragged2e}
\usepackage[utf8]{inputenc}
\usepackage[english]{babel}

\usepackage{natbib}
\usepackage{authblk}
\usepackage{float}

\title{Transitioning from A-level to higher education physics courses: where are the knowledge gaps?}
\author[1]{Paul Mackay}
\author[2]{Oto-obong Inyang}
\affil[1]{Department of Physics, Durham University}
\affil[2]{Department of Physics, Durham University}
\date{September 2026}

\begin{document}

\maketitle
\begin{abstract}

Historical post-16 data for UK A-Level Physics, Maths and Further Maths are analysed to determine the relative popularity of A-Level exam boards. A curriculum mapping of the exam board specifications is implemented and  compared against first year syllabi of undergraduate physics courses at high-ranked UK universities. Gaps in curricula are identified and recommendations are made to both university module designers and A-level exam board curriculum setters, with regards to remedial action and a reconsideration of included content. 

\end{abstract}

\section{Introduction}
The success and retention of physics students in higher education (HE) can be linked to their ability to successfully transition from school curricula to undergraduate (UG) level courses (\cite{PorterChuIvie2024, ODonnellKeanStevens2016}). There are two main approaches used in deciding what content to include in first year UG courses: an accelerated model where students see university-level material early, with remedial support (\cite{capece2024}); and a consolidated approach (see the spiral progression approach of (\cite{kampf2025}). In the UK, the QAA also highlights progression as a key component of a physics degree (\cite{QAA_PhysicsBenchmark_2025}). 
Mathematical demand of HE is also a key factor in student retention (\cite{StewartHansenBurkholder2022}) with mathematics typically prerequisite for a Physics degree. It is a powerful tool and language in understanding Physics concepts and problem solving, which demands from students the quality of bending physical interpretations to mathematical formalism(\cite{redish2015language, redish2005}). This has been described by Redish as bilingual use of mathematics (\cite{redish2015language}).  

Non-continuation rates show that nationally, around 6\% of students do not continue past their first year of study (\cite{HESAcontinuation}) and we consider reasons for this. One motivator is the development of expertlike views (\cite{Bates2011Attitudes}) - do some students fail to develop such views by learning content perceived too trivial and feel they are not becoming experts? Elsewhere, course content is another factor that affects student retention (\cite{Sneddon2022}), with participants in that study variously saying content ramped up exponentially or was too overlapping of school curricula (see also \cite{ThompsonPyper2006WhyPhysics}).

Similarly, in a study of female students transitioning from GCSE to A-Level (\cite{CassidyCattanCrawfordDytham2018}), one complaint was on the lack of depth (as opposed to breadth) in curricula with too much time dedicated to older material. We posit that the same could be true from A-Level to degree.

In this paper, we perform a curriculum mapping (\cite{Harden2001}) of UK Physics,  Maths and Further Maths A-Level exam boards, to highlight differences in content. Alongside statistics on numbers of candidates taking each qualification, we arrive at values for proportions of students in a typical undergraduate physics cohort having particular knowledge gaps. Where close to 100\% of students have studied a topic is not of interest (though a spiral curriculum model may still be worthwhile as a study in biosciences (\cite{Jones2015}) showed that even high ability A-Level students typically forgot a huge amount of content on reaching university); at the other end of the scale, topics that almost no students have seen before are obvious targets for university teaching, and can be considered a quirk of that A-Level exam board. However, for key topics (and these we determine by reviewing the common topics taught at first year in top UK universities), where a moderate number of students (but certainly not close to all) have seen a topic at A-level, such that a significant group of students will be seeing content for the first time, while another significant group will not, then we consider this a cohort knowledge gap and we make recommendations. The issue for A Level Further Maths is more significant, as some students will not have taken the qualification at all. Six($6$) out of forty($40$) Institute of Physics(IOP) accredited universities suggest Further Maths (or Chemistry) as the third A-level entry qualification to a Physics degree (\cite{Fox2025}).

Furthermore, undergraduate curricula may have been static for many years, unreactive to A-level curricula changes, with presupposition of certain content being taught. For example, previously Edexcel A-Level Maths included a Mechanics module with content on moment of inertia calculations and other rotational mechanics (for example, see the exam paper: \cite{Edexcel_ALevel_Maths_2013}), but this material is no longer covered in A-Level Maths (by any board). A majority of students will not have seen this material yet university teaching staff may be unaware.

The practical content also differs significantly between exam boards, such that undergraduates will have differing experiences in quantity and coverage of experiments, and use of different apparatus. However, this is beyond the scope of this paper. 

Furthermore, the focus of this paper is largely on students in England, Wales and Northern Ireland, as in Scotland (approximately 2\% of all UK students), most students take Scottish Highers or Advanced Highers(\cite{SQA_Higher_Physics_2026, SQA_AdvancedHigher_Physics_2026}). Advanced Highers cover additional topics typical of first year UG courses, with most Scottish university courses one year longer than equivalent courses elsewhere in the UK. Consequently, they offer accelerated entry onto Scottish courses with sufficient grades; however, reduced content of Highers alone may cause significant disparity for Scottish students attending other UK universities (having missed large parts of what is core material offered elsewhere (e.g. circular motion, thermal physics, quantum (beyond photoelectricity), fields calculations). Scottish schools can offer A-levels but it is rarely done (largely the domain of independent fee-paying schools (\cite{SchoolAtlas_IndependentSchoolsScotland})). 

Elsewhere, international students may study a wide range of curricula in their home countries, with varying content, but this is also beyond the scope of this paper (and of limited use given the different demographics at any one university).

\section{Methodology}
Data on numbers of candidates entered nationally for A-level Maths, Further Maths and Physics for each of the most common exam boards were averaged over five years to determine typical percentages of students who have studied with each exam board. For comparison, institutional data of A-level exam board choice for a top-ranked Russell Group university was analysed over a similar timescale, to see that the distributions of A-level choices of physics undergraduates was broadly similar to the national picture, such that there is confidence that the national figures can be taken as a proxy for a typical university demographic. This also largely reflects the reasonable assumption that students have little to no choice over exam boards used by their schools. Error bars on the data were calculated based on  UCAS data being rounded to the nearest 5\% in addition to standard error calculation. 

A curriculum mapping was performed, whereby specifications for each exam board, for each of the three subjects, were analysed to identify topics taught. This was done at the finest level of individual specification points (e.g. "Moment of force about a point" as opposed to "Mechanics"). Typically, this involved breaking down each specification into hundreds of topics. Author judgement was used to equate topics in cases where they were described differently by various boards. 
Combining the curriculum mapping with the typical percentages having studied with each board, it was possible to determine, for each topic, a percentage of UG physics students who will (or will not) have seen it at A-level. Students without A-Levels are not considered in this analysis, as these will largely consist of Scottish or international students, the numbers of which will vary more widely between institutions, based on factors such as reputation and geographical location. 
Publicly available descriptions of module content from a selection of high-ranking UK universities were then used to identify common first year topics, to identify topics: being seen by most students for the first time at UG; that most students had seen before at A-level; or a more mixed picture.

\section{Comparison of A-Level Physics exam boards}
The post-16 landscape for physics education is varied across the UK. 

In England, Wales and NI, schools have access to an open qualification market, and can deliver courses offered by AQA, OCR, Edexcel, Eduqas or CCEA (\cite{AQA_Physics_7408_2026,Edexcel_Physics_9PH0_2026,Eduqas_Physics_A730QS_2026,CCEA_Physics_2016_2026, OCR_PhysicsA_H556_2026}).

\begin{figure}[H]
  \includegraphics[width=\linewidth]
  {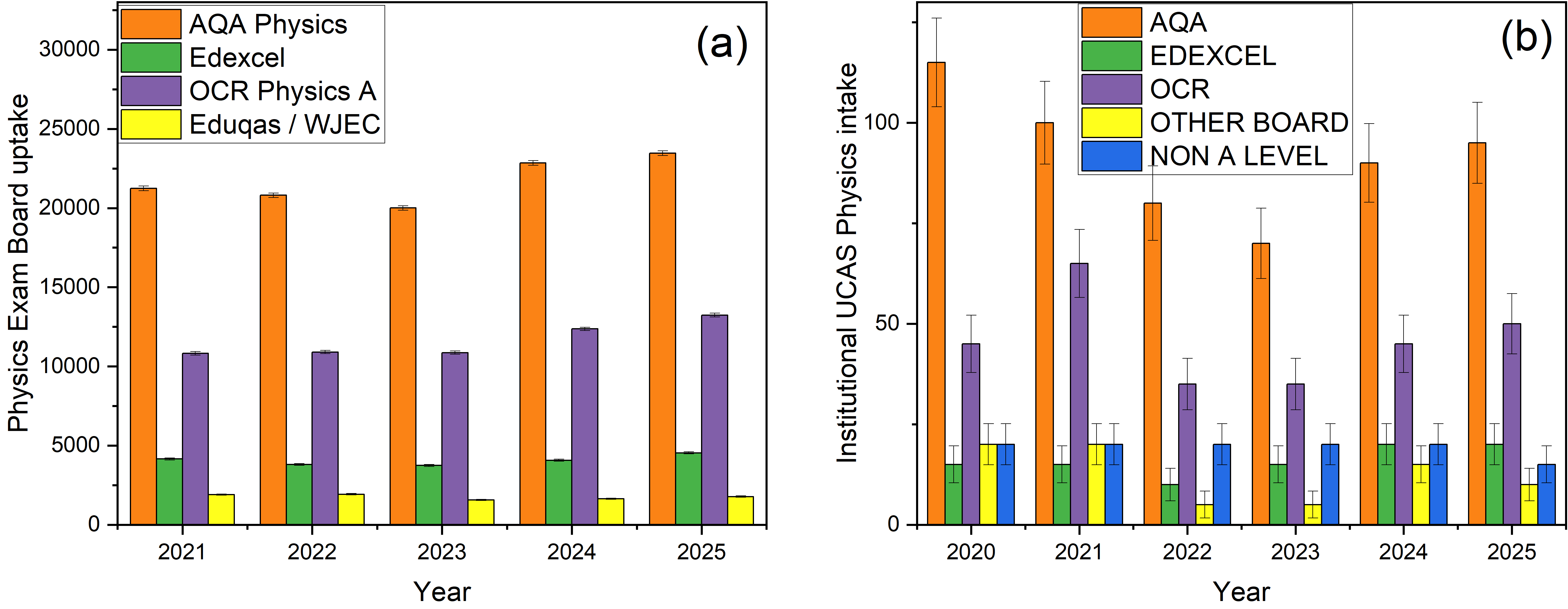}
  \centering
  	\caption{A-Level Physics exam board (a) uptake (b) institutional UCAS intake over time. "Other board" includes Eduqas and CCEA and "Non A-Level" includes Scottish Highers and any international boards. (\cite{UCAS_FOI})}
  \label{fig:UCAS_physics}
\end{figure}

For A-Level Physics, the most popular exam board has consistently been AQA (\cite{UCAS_FOI}), representing roughly half of all entries (Figure \ref{fig:UCAS_physics}). 

AQA is considered to contain the most standard question style, with students applying knowledge to a variety of contexts. AQA contains an optional module in the second year of study, presenting challenges to creators of first year undergraduate courses, with regards to prior knowledge expectations.

The AQA optional module covers either Astrophysics, Medical Physics, Engineering Physics, Turning Points in Physics, or Electronics. Of these, Astrophysics is the most popular, corresponding to 54\% of AQA entries, approximately 26\% of all physics entries. One consequence is that AQA candidates choosing any option other than astrophysics (28\% of all candidates) will not study space science at A-Level. This may be significant, as studying space physics is often linked to student engagement \cite{OsborneCollins2000}. Students are also often told that the space sector is one of the fastest growing \cite{UKSpaceAgency_CareersLeaflet} (though that may not reflect reality (\cite{PwC_SpaceSector})) with \cite{Sneddon2022} espousing career availability as major motivator for physics students. 
Turning Points (32\% of AQA students, 17\% overall) is the main source of UG students who may have seen time dilation formula before.
Only 15\% of AQA entries (8\% overall) are Engineering Physics. This is significant, largely accounts for overall 91\% of UG students who have not previously seen advanced rotational motion concepts such as moment of inertia (this topic is not covered in the core of any other specification, nor is it encountered in Maths or Further Maths A-level current specifications 
A note can be made here on AQA students who take neither Astrophysics nor Medical Physics options (26\% of all physics students), the options where lenses are described. While refraction generally is part of core material at both GCSE and A-Level, it is possible for students to begin a university physics course without having ever studied lenses (which appears at GCSE only for students taking Physics separately (also known as Triple science when taken with individual Biology and Chemistry GCSEs, too) as opposed to a combined science (dual award) qualification - a number which is also decreasing \cite{FFTDatalab_GCSEResults2026}) as there was a 6\% decrease from 2024 to 2025 in Separate Physics entries. Furthermore, up to 10\% of UK schools do not offer separate Physics GCSE.

After AQA, OCR(A) was the next most popular board (taken by 29\% of all candidates). OCR questions are typically more contextual. 
A key topic covered by OCR(A) - and missing from AQA - is Archimedes' principle. Unlike AQA, OCR(A) does not contain optional modules, therefore all students see the same content - with the same gaps. 

Edexcel was less popular for Physics (10\% overall), with a reputation of being more analytical, with questions requiring use of data. 

In addition to these, Eduqas (a Welsh exam board, also known as WJEC) or CEAA (a Northern Irish exam board)  were taken by smaller numbers, with even fewer again for OCR(B) (a more contextual variant of OCR(A) but soon to be phased out). Eduqas questions are very straightforward, with little context, though notably the syllabus does include a few topics not covered by other boards, such as lasers and Ampere's Law. However, in light of such small numbers ($<$5 \% in each case), we will not focus on these boards, except to include in the overall percentage calculations. 

Note that for England/Wales/NI, there is no requirement to choose UK A-level boards. Indeed, there has been a gradual rise in students taking International A-Level physics. Some suggested reasons for this are as follows: they retain the modular structure that UK A-level exam boards had pre-2015 (\cite{DfE2018ASAlevelChanges}) allowing for frequent exam sittings and resits, with students needing to retain less knowledge at any one time (cf. terminal exams); the qualification is nominally academically equivalent yet language is simplified so may be preferable in regions with a large number of students for whom English is not first language, which often results in context being stripped back; they contain no practical element so are preferable for home schooled or private individuals. Regardless, it is useful to note an increasing number of undergraduates who may lack practical awareness, who have never sat terminal exams for content covering an entire course, and for whom context may have been toned down.

\subsection{Typical first year Physics content of high-ranked UK universities}
To ascertain what a typical physics curriculum of a high-ranked university (\cite{CompleteUniversityGuide2027}) contains, publicly available module descriptions were analysed. 

Of course, there is a lot of the commonality between content for two main reasons: to be accredited by the IOP, physics degrees previously required certain content, though they did not specify at which stage those items were taught (\cite{IOP_PhysicsDegree_2011}) - this is now outdated, as the accredidation framework (\cite{IOP_AccreditationFramework_2022}) merely stipulates the percentage of physics content, but does not specify what that content is; secondly, many institutions use a general text like \cite{YoungFreedman2023} or \cite{HallidayResnickWalker2018} as a loose framework for first year content. 

Classical mechanics, oscillations, waves and optics, quantum mechanics, relativity, circuits and electric and magnetic fields are seen to be core material in all cases, and within each of those topics, broadly the same concepts are covered. Thermal physics is also studied as core material by a majority of institutions. 

In our sample, only 30\% of institutions offered astrophysics as a core component, though it is usually offered optionally. Astrophysics content varies significantly, however, with some offering more observational astronomy, and others largely theoretical. Similarly, particle and nuclear physics is core material for a minority of institutions in our sample.

Therefore, we define the following topics as typical core content of a first year UG degree: classical mechanics, oscillations, waves and optics, quantum mechanics, relativity, circuits, electric and magnetic fields and thermal physics.

\subsection{Identifying the gaps}
We now identify which typical core content are not commonly taught at A-Level (see Table \ref{tab:unseentopics}). Of most interest for remedial action are those topics around the 25-75\% mark, where reteaching to the whole cohort may be perceived negatively by students with prior exposure; whereas if a majority (say, over 90\%) of students have not seen a topic, then it makes sense to simply teach that. Using exam board data, we can determine a percentage of students unlikely to have previously studied particular concepts common to first year UG physics courses from a range of top UK universities (\cite{Durham_Physics_Year1_2026,Oxford_Physics_FirstYear_2026,Warwick_Physics_FirstYearModules_2026,Imperial_Physics_Year1Modules_2026,Manchester_Physics_BSc_2026, Bath_Physics_BSc_2026,Bristol_Physics_BSc_2026,Edinburgh_Physics_DPT_2026,StAndrews_Physics_ProgrammeRequirements_2026,Birmingham_Physics_BSc_2026}).

\begin{table}
    \centering
    \begin{tabular}{cc}
    \toprule
       \textbf{Topic}  & \textbf{\% not seen } \\
       \hline
        Stefan-Boltzmann Law & 32  \\
        Dispersion & 40  \\
        Archimedes' Principle & 59  \\
        Lenses & 61  \\
        Relativity & 72  \\
        Huygen's Principle & 72  \\
        Stoke's Law & 72  \\
        Rotational motion (beyond basic circular motion) & 91  \\
        Ampere's Law & 96  \\
        Bernoulli & 96  \\
        Inductors & 99 \\
    \bottomrule
    \end{tabular}
    \caption{Table shows common topics that appear in first year undergraduate physics courses, alongside a calculated percentage of students who will not have encountered it at A-Level. Values are calculated for physics students who studied A-Level physics with AQA, OCR(A) and Edexcel, which together typically account for approximately 95\% of students.}
    \label{tab:unseentopics}
\end{table}

\subsection{Recommendations (Physics)}
Based on the above analysis, we identify two main issues. 
Firstly, that A-level exam board specifications contain too much content. In particular, there are numerous topics covered at A-Level, which are generally not expanded on in most undergraduate physics courses, suggesting that universities do not consider these topics important. In particular, particle and nuclear physics are not usually revisited until second or third year. Given that students are known to have difficulty recalling content after a single year, the issue is likely to be compounded after two or three years before revisiting those topics.

Conversely, university courses could consider excising some of the most basic concepts in kinematics and electric circuits. Furthermore, while core first year content is broadly similar between providers, we suggest further increasing homogeneity in first year content between course providers with respect to alignment with A-level specifications as discussed.

Otherwise, where the course structure is inflexible, self-study materials at an A-Level standard with self-assessment questions and in accessible formats should be made available for those topics identified as being previously unseen by a significant proportion (again, around 25-75\% is suggested).

Taken altogether, these recommendations should ease the transitions from A-Level to university, with positive outcomes for student retention and progress. 

\section{Comparison of Mathematics and Further Mathematics A-Level exam boards}
\begin{figure} [H]
  \includegraphics[width=\linewidth]
  {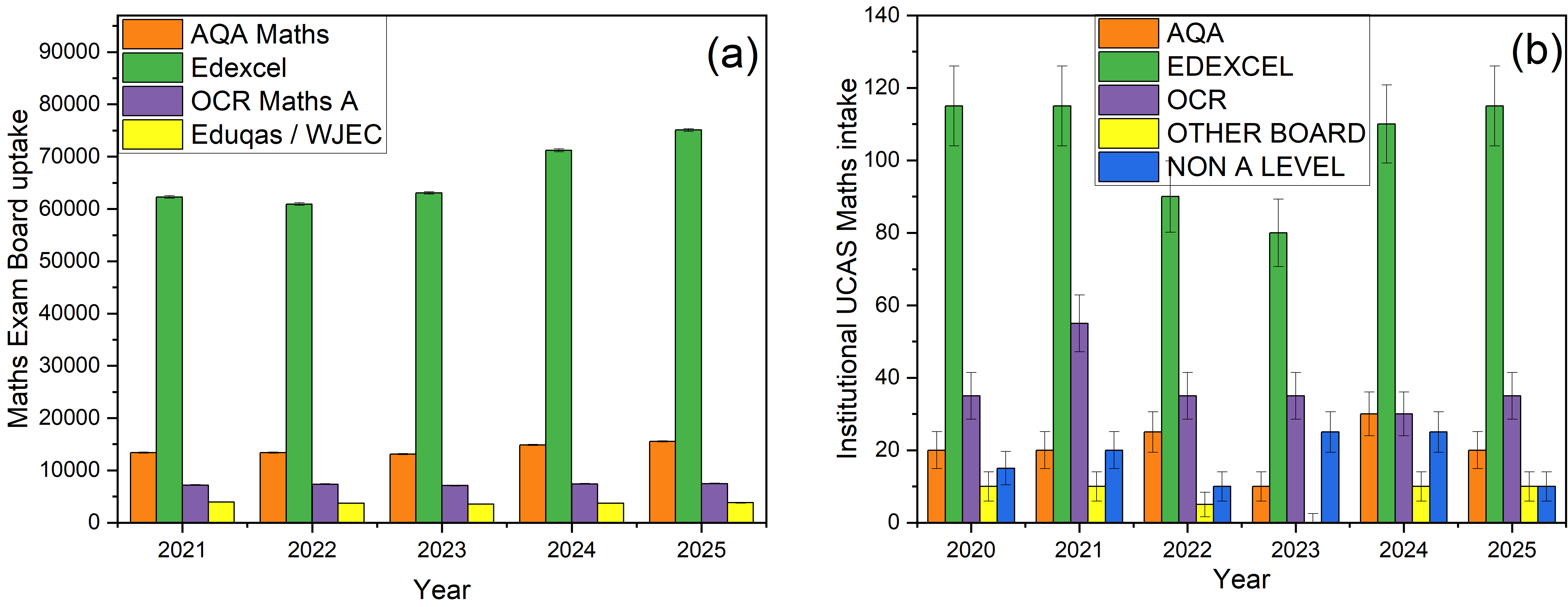}
  \centering
  	\caption{A-Level exam board choices between 2021 to 2025 for (a) Mathematics and (b) Further Mathematics. Similar to Physics, "Other board" includes Eduqas and CCEA, and "Non A-Level" includes Scottish Highers and any international boards. (\cite{UCAS_FOI})}
  \label{fig:UCAS_Maths}
\end{figure}

The awarding bodies offering A-level Mathematics and Further Mathematics specifications in the United Kingdom are the AQA, Edexcel, OCR and Educas/WJEC. According to data on the uptake of each exam board, represented in Figure \ref{fig:UCAS_Maths}a, Edexcel is overwhelmingly the dominant board for A-level Mathematics nationally. Its entries are substantially higher than all the other boards from $2021$ to $2025$. The average percentage documented number of intake over the 5 years is about $68\%$ with a 5\% increment from 2021 to 2025, representing a clear upward trajectory(See Figure \ref{fig:UCAS_Maths} (a)). AQA Mathematics entries, whilst the second largest from the board intake data, account for about 14\% of entries, are considerably smaller and show modest and consistent growth similar to OCR in the institutional data represented in Figure \ref{fig:UCAS_Maths}b. The other boards account for a comparatively smaller share of the national Mathematics entries, with figures below 8\%. This shows a significant gap between Edexcel and the other boards, which is also reflected in the entry-level undergraduate student cohort(see Figure  \ref{fig:UCAS_Maths}b). Whilst Edexcel is the dominant board represented in most high-ranking institutions in the UK, there is diversity in the percentage representation of the other boards across institutions.  

The content mapping across AQA, Edexcel, OCR and Educas/WJEC reveals a broadly consistent A-level Mathematics core curriculum underpinned by a shared national framework(\cite{AQA_Maths_7357_2026,Edexcel_ALevel_Maths_2013,OCR_MathsA_H240_2026,Eduqua_WJEC_MathsA_2026}). Topics covered include algebra and functions, coordinate geometry, trigonometry, exponential and logarithms, differentiation, integration, sequences and series, vectors, probability, statistics and mechanics. This convergence reflects the regulatory oversight of the Office of Qualifications and Examinations Regulation(Ofqual), which mandates a common content baseline across all approved A-level specifications.

\begin{figure}[H]
  \includegraphics[width=\linewidth]
  {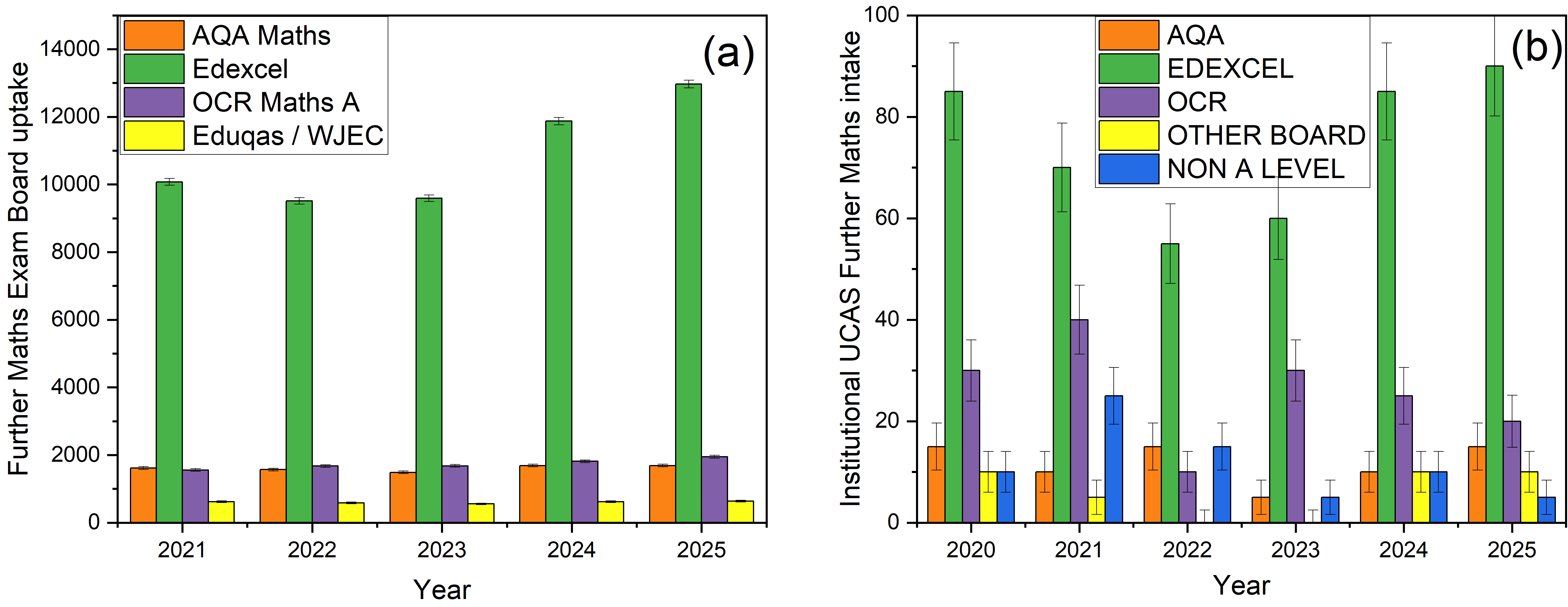}
  \centering
  	\caption{A-Level exam board choices between 2021 to 2025 for (a) Mathematics and (b) Further Mathematics. Similar to Physics, "Other board" includes Eduqas and CCEA, and "Non A-Level" includes Scottish Highers and any international boards \cite{UCAS_FOI}.}
  \label{fig:UCAS_FMaths}
\end{figure}

Similarly, the national pattern for Further Mathematics closely mirrors that of Mathematics, with Edexcel the dominant exam board as shown in Figure \ref{fig:UCAS_FMaths}. Further Mathematics uptake is substantially lower than Mathematics across every board. Even Edexcel, the largest provider, records entries roughly five times lower than its Mathematics cohort, underscoring the limited reach of Further Mathematics provision nationally. Also, while AQA is the second largest provider in Mathematics, its Further Mathematics entries are smaller, suggesting that many AQA-centre schools do not extend to offer Further Mathematics. This is consistent with the known challenge of specialist teacher availability.

The Further Mathematics landscape introduces considerably more variability\cite{AQA_FurtherMaths_7367_2026,Edexcel_ALevel_FurtherMaths_2013,OCR_FMathsA_H245_2026,Eduqua_WJEC_FMathsA_2026}. All boards share a compulsory core, accompanied by optional components. This optionality means that even students who hold a Further Mathematics qualification may not have studied all relevant topics. For example, a student who chooses a Statistics option may have no exposure to advance mechanics content such as simple harmonic motion, damped oscillations or circular motion, all of which are directly relevant to Physics.

\subsection{Typical first year Mathematics content of high-ranked UK universities}
Following a similar approach to the Physics content analysis above, a review of publicly available module information from mathematics-related modules within Physics degree programmes was conducted across five high-ranking UK universities, as ranked by the Complete University Guide (\cite{CompleteUniversityGuide2027}). This analysis was undertaken to establish a representative picture of the mathematical content typically encountered by students in the first year of an undergraduate Physics degree.

The five universities examined all deliver substantial first-year mathematical provision within their Physics programmes. Although each institution structures and sequences its content differently, there is a striking degree of convergence around a shared core curriculum, alongside notable differences in depth, breadth and disciplinary emphasis. Complex numbers, differential and integral calculus, Linear algebra, vectors and vector algebra and ordinary differential equations(ODEs) were found to feature across all five institutions. Complex numbers was found to be a foundational topic with coverage consistently encompassing the Argand diagram, modulus and argument, De Moivre's theorem and exponential form. This uniformity reflects the centrality of complex numbers to quantum mechanics, Optics, and alternating current (AC) circuit analysis encountered in parallel Physics modules. In addition, it addresses the gap identified among students with no A-level Further Mathematics experience. However, it would require deliberate scaffolding to better support this student category. Ordinary differential equations content was often explicitly contextualised within Physics applications such as oscillations, damping and resonance.

Despite shared foundations among institutions, they differ meaningfully in scope and depth, with some institutions integrating experimental data analysis distinctively. Others stand out in their early and systematic treatment of Fourier series, which appear formally later in the year. Some also place notable emphasis on probability and statistics, reflecting a deliberate effort to help students develop the skill of connecting mathematical tools directly to Physical theory from the outset, while another incorporated an introduction to group theory, reflecting its Theoretical Physics tradition and importance of symmetry in advance Physics.

Overall, institutional mathematics content structure and sequences reflected each institution's broader pedagogical philosophy and the degree to which mathematical content is integrated with or separated from the Physics delivery. 

\subsection{Identifying the gaps}

The standard Mathematics entry requirement for most Physics undergraduate programmes is A-level Mathematics, with A-level Further Mathematics recognised as a valuable additional qualification. However, the availability of Further Mathematics at A-level remains uneven across the country, largely due to a national shortage of specialist Mathematics teachers, which limits the number of schools able to offer it (\cite{straker1988mathematics,bullock1992teacher,sani2022longitudinal}). As a result, a significant proportion of students arrive at the university without exposure to certain advanced mathematical concepts, creating knowledge gaps that can affect the student's experience during this critical period.
\begin{table}[!ht]
\caption{Key advanced mathematical concepts and their associated area of Physics where they are relevant.}
\centering
\begin{tabular}{ |p{2.9cm}||p{9cm}|  }
\toprule
 \hline
 \textbf{Advanced Mathematical topics}& \textbf{Physics topics}\\
 \hline
 \hline
 Complex Numbers   & Quantum mechanics, AC circuits, Optics, Oscillations, Waves    \\
 \hline
 Matrices&   Quantum Mechanics, Special relativity, Optics, Oscillations\\
 \hline
Taylor series &Classical mechanics, Special relativity, thermodynamics, Electromagnetism, Quantum mechanics \\
 \hline
 \bottomrule
 \end{tabular}
 \label{tab:F.Maths_to_Physics_Topics}
\end{table}
An analysis based on mapping of individual examination board specifications and subject content reveals that students without A-level Further Mathematics typically have had no prior engagement with key mathematical topics including Complex Numbers, Matrices, and Taylor series. These are not peripheral areas of mathematics, but underpin fundamental concepts across several core areas of Physics, as outlined in Table \ref{tab:F.Maths_to_Physics_Topics}. This points strongly to the need for diagnostics assessment at entry and flexible bridging support tailored to the diverse mathematical background of the incoming cohort.

\subsection{Recommendations (Maths)}

To address the mathematical knowledge gap at the transition to undergraduate Physics, a coordinated set of interventions is recommended. These may begin with a mandatory diagnostic assessment at the entry to establish each student's mathematical background, which could be accompanied by freely accessible online bridging resources covering the highest-priority topics. Early addressing of the high-priority topics in the first year will provide the students with the necessary tools to better understand the Physics content. Differentiation of workshops and tutorial provisions to cater to students with varying levels of preparedness, offering consolidation and extension opportunities to students. 

Equally important is the embedding of contextual mathematical support within the first year Physics modules rather than treating mathematics as a standalone remediation activity. This would improve student understanding and motivation. Alongside this, a proactive engagement of institutions with schools and colleges to raise awareness of the value of Further Mathematics for Physics progression, particularly in settings where specialist teacher shortage limits its availability. Collectively, these measures are expected to reduce the transition gap meaningfully and improve student experience and academic outcomes in the first year of undergraduate study.

\bibliography{references}

\end{document}